\documentclass[a4paper]{jpconf}
\usepackage{graphicx}
\usepackage{amsmath}
\usepackage{amsfonts}
\begin{document}
\title{Geometry of surfaces associated to grassmannian sigma models}

\author{L Delisle$^1$, V Hussin$^{2,3}$ and W J Zakrzewski$^4$}

\address{$^1$Institut de math\'ematiques de Jussieu--Paris Rive Gauche, UP7D--Campus des Grands Moulins, B\^{a}timent Sophie Germain, Cases 7012, 75205 Paris Cedex 13.}
\address{$^{2}$D\'epartement de math\'ematiques et de statistique, Universit\'e de Montr\'eal, C. P. 6128, Succ. Centre-ville, Montr\'eal (Qu\'ebec) H3C 3J7, Canada.}
\address{$^3$ Centre de recherches math\'ematiques, Universit\'e de Montr\'eal, C. P. 6128, Succ. Centre-ville, Montr\'eal (Qu\'ebec) H3C 3J7, Canada.}
\address{$^4$ Department of mathematical sciences, University of Durham, Durham DH1 3LE, United Kingdom.}

\ead{laurent.delisle@imj-prg.fr, hussin@dms.umontreal.ca, w.j.zakrzewski@durham.ac.uk}

\begin{abstract}
We investigate the geometric characteristics of constant gaussian curvature surfaces obtained from solutions of the $G(m,n)$ sigma model. Most of these solutions are related to the Veronese sequence. We show that we can distinguish surfaces with the same gaussian curvature using additional quantities like the topological charge and the mean curvature. The cases of $G(1,n)=\mathbb{C}P^{n-1}$ and $G(2,n)$ are used to illustrate these characteristics.
\end{abstract}

\section{Introduction}
In recent papers [1,2], we have classified some relevant solutions of the grassmannian $G(m,n)$ sigma model that are associated to constant gaussian curvature surfaces in $su(n)$. In our construction we have found, among others, some non-equivalent solutions with the same constant gaussian curvature. In the non-holomorphic case [2], we have considered a particular set of solutions starting from the knowledge of the corresponding solutions of $G(1,n)=\mathbb{C}P^{n-1}$, the so-called Veronese solutions. In the holomorphic case [1], we presented some conjectures and constructed some solutions which are not related to the Veronese ones. These results clarified and extended results obtained elsewhere [3-7].

In this contribution, we aim to show that some of the surfaces that have the same constant gaussian curvature and correspond to non-equivalent solutions of $G(m,n)$ up to a gauge transformation, may be distinguished by other geometric characteristics such as the topological charge and the mean curvature. The case of $G(2,n)$ will be discussed in detail to show how this works out.

In Section 2, we discuss the $G(m,n)$ sigma model and we define the geometric quantities for the surfaces associated to $G(m,n)$. In particular, we recall a class of solutions of the model that lead to surfaces with constant gaussian curvature and explain the relation with the Veronese sequence. We compute explicitly the additional geometric characteristics for these surfaces. In Section 3, we show how these quantities could exhibit the differences between the surfaces with same constant gaussian curvature. We re-visit the case of $G(1,n)=\mathbb{C}P^{n-1}$ and give some general results for $G(m,n)$. We show, in the case of $G(2,n)$, how different solutions with the same gaussian curvature may be distinguished by calculating their topological charge and/or their mean curvature. We conclude this section with the case of non-Veronese holomorphic solutions. Section 4 presents our conclusions and mentions our future plans.

\section{Surfaces associated to solutions of $G(m,n)$}

\subsection{The model}
The two-dimensional $G(m,n)$ sigma model is a field theory [8] defined on the complex plane $\mathbb{C}$ which has as a target space the grassmannian manifold $G(m,n)$ :
\begin{equation}
G(m,n)\cong \frac{U(n)}{U(m)\times U(n-m)},\quad n>m,\label{Gmn}
\end{equation}
where $U(k)$ is the set of $k\times k$ unitary matrices. The field $Z(x_+,x_-)$ defined on an open and simply connected subset $\Omega$ of ${\mathbb C}$ thus takes values in $G(m,n)$. The elements $Z$ of $G(m,n)$ are parametrized by $n\times m$ matrices and satisfy $Z^{\dagger}Z=\mathbb{I}_m$. Moreover,
 they correspond to critical points of the energy functional defined via the Lagrangian density
\begin{equation}
\mathcal{L}(Z)=\frac{1}{2}\hbox{Tr}\left[(D_+Z)^{\dagger}D_+Z+(D_-Z)^{\dagger}D_-Z\right],\label{Lagrangian}
\end{equation}
where $D_{\pm}\Lambda=\partial_{\pm}\Lambda-\Lambda (Z^{\dagger}\partial_{\pm}Z)$ are the covariant derivatives, $\partial_{\pm}=\partial_{x_{\pm}}$ and $(x_+,x_-)$ are complex local coordinates on $\Omega$. We consider the case of
 $\Omega=\mathbb{C}$ and require the energy of these fields to be finite.
 In order for this to be the case we impose the boundary conditions
$D_{\pm}Z\longrightarrow 0$ as $\vert x\vert\longrightarrow\infty$.
With such boundary conditions, the complex plane $\mathbb{C}$ is compactified into the two-sphere $S^2$ via the stereographic projection and, as a consequence, the fields $Z$ are harmonic maps [9] from $S^2$ into the Grassmann manifold $G(m,n)$. 

Using the variation of the energy, we deduce the Euler-Lagrange equations of the model given as ($\pm\longleftrightarrow\mp$)
\begin{equation}
D_+D_-Z+Z(D_-Z)^{\dagger}D_-Z=0,\quad Z^{\dagger}Z=\mathbb{I}_m.\label{ELeq}
\end{equation}
The finite energy solutions of these equations are fully known in the $G(1,n)\cong\mathbb{C}P^{n-1}$ case [9]. They are given by
\begin{equation}
Z_i=\frac{P_+^if}{\vert P_+^if\vert},\quad i=0,1,\cdots,n-1,\label{solG1n}
\end{equation}
where $f=f(x_+)\in\mathbb{C}^n$ is holomorphic and $P_+$ is an orthogonalizing operator defined recursively as
\begin{equation}
P_+^0f=f,\quad P_+f=\partial_+f-\frac{f^{\dagger}\partial_+f}{\vert f\vert^2}f,\quad P_+^if=P_+(P_+^{i-1}f),\quad P_+^nf=0.
\end{equation}

For the $G(m,n)$ model with $m\geq 2$, the complete set of solutions is not known, in an explicit way, but we can use the solutions (\ref{solG1n}) of the $G(1,n)$ model to construct  particular classes of them:
\begin{equation}
Z_{(i_1,i_2,\cdots,i_m)}=\left(\frac{P_+^{i_1}f}{\vert P_+^{i_1}f\vert},\frac{P_+^{i_2}f}{\vert P_+^{i_2}f\vert},\cdots,\frac{P_+^{i_{m}}f}{\vert P_+^{i_m}f\vert}\right),\quad 0\leq i_1<i_2<\cdots<i_m\leq n-1.
\label{zmn}
\end{equation}

A convenient way to reformulate the model in a gauge-invariant way involves using orthogonal projectors [8]. Indeed, for $G(m,n)$, we define a rank $m$ hermitian orthogonal projector $\mathbb{P}$ as
\begin{equation}
\mathbb{P}=ZZ^{\dagger}.
\end{equation}
This projector satisfies
\begin{equation}
\mathbb{P}^2=\mathbb{P}^{\dagger}=\mathbb{P},\quad \hbox{Tr}(\mathbb{P})=m.
\end{equation}
The Lagrangian density (\ref{Lagrangian}) and the Euler-Lagrange equations (\ref {ELeq}) are written in an equivalent way as
\begin{equation}
\mathcal{L}(\mathbb{P})=\frac12\hbox{Tr}(\partial_+\mathbb{P}\partial_-\mathbb{P}),\quad [\partial_+\partial_-\mathbb{P},\mathbb{P}]=0,\quad \mathbb{P}^2=\mathbb{P}.\label{ELproj}
\end{equation}

A solution of the type (\ref{zmn}) leads to a projector of the form:
\begin{equation}
\mathbb{P}_{\beta}=\sum_{j=0}^{n-1}\beta_j\, \mathbb{P}_j,\quad \mathbb{P}_j=Z_jZ_j^{\dagger}=\frac{P_+^jf(P_+^jf)^{\dagger}}{\vert P_+^jf\vert^2},\label{solGmn}
\end{equation}
where $\beta$ is a $n-$column vector such that $\beta_j=0$  or 1 for all $j$ and $\sum_{j=0}^{n-1}\beta_j=m$. 

The key, in constructing surfaces from the solutions of the $G(m,n)$ model, is to observe that the Euler-Lagrange equations (\ref{ELproj}) may be written as a conservation law:
\begin{equation}
\partial_+\mathbf{L}-\partial_-\mathbf{L}^{\dagger}=0,\quad \mathbf{L}=[\partial_-\mathbb{P},\mathbb{P}].
\end{equation}
Then, using the Poincar\'e lemma and the fact that $\Omega$ is simply connected [10], we may define a surface $\mathbf{X}\in su(n)$ (the set of $n\times n$ hermitian and traceless matrices) via its tangent space as
\begin{equation}
d\mathbf{X}=\mathbf{L}^{\dagger} dx_++\mathbf{L}dx_-\label{defsurfaces}
\end{equation}
or explicitly
\begin{equation}
\partial_+\mathbf{X}=[\partial_+\mathbb{P},\mathbb{P}],\quad \partial_-\mathbf{X}=-[\partial_-\mathbb{P},\mathbb{P}].
\label{derX}
\end{equation}

\subsection{Lagrangian and topological densities}

Let us recall that the topological density is defined by [8] 
\begin{equation}
\mathcal{Q}(\mathbb{P})=\frac{1}{2}\hbox{Tr}\left[(D_+Z)^{\dagger}D_+Z-(D_-Z)^{\dagger}D_-Z\right]=\frac12\hbox{Tr}\left[\mathbb{P}[\partial_-\mathbb{P},\partial_+\mathbb{P}]\right],\label{topodens}
\end{equation}
showing, in particular, that for holomorphic (or anti-holomorphic) solutions, which satisfy $D_-Z=0$ (or $D_+Z=0$), it coincides with the Lagrangian density (up to a sign).

Hence, for the solutions (\ref{solGmn}), we get the explicit expressions for the Lagrangian density (\ref{ELproj}) and the topological density (\ref{topodens}) as follows:
\begin{equation}
\mathcal{L}(\mathbb{P}_{\beta})=\frac12\sum_{j=1}^{n-1}(\beta_{j-1}-\beta_j)^2\frac{\vert P_+^jf\vert^2}{\vert P_+^{j-1}f\vert^2},\quad \mathcal{Q}(\mathbb{P}_{\beta})=\frac12\sum_{j-1}^{n-1}(\beta_{j-1}-\beta_j)\frac{\vert P_+^jf\vert^2}{\vert P_+^{j-1}f\vert^2}.
\end{equation}
Let us exhibit some properties of the topological charge. We may rewrite the topological density as 
\begin{equation}
\mathcal{Q}(\mathbb{P}_{\beta})=\mathcal{Q}\left(\sum_{j=0}^{n-1}\beta_j\mathbb{P}_j\right)=\frac{1}{2}\sum_{j=0}^{n-1}\beta_j\left(\frac{\vert P_+^{j+1}f\vert^2}{\vert P_+^jf\vert^2}-\frac{\vert P_+^jf\vert^2}{\vert P_+^{j-1}f\vert^2}\right)=\sum_{j=0}^{n-1}\beta_j\mathcal{Q}(\mathbb{P}_j),
\end{equation}
showing that it is a purely additive quantity. Furthermore, using the topological property
\begin{equation}
\mathcal{Q}(\mathbb{P}_j)=\frac{\vert P_+^{j+1}f\vert^2}{\vert P_+^jf\vert^2}-\frac{\vert P_+^jf\vert^2}{\vert P_+^{j-1}f\vert^2}=\partial_+\partial_-\ln(\vert P_+^jf\vert^2),
\end{equation}
the topological density takes the compact form
\begin{equation}
\mathcal{Q}(\mathbb{P}_{\beta})=\partial_+\partial_-\ln\prod_{j=0}^{n-1}\vert P_+^jf\vert^{\beta_j}.
\end{equation}

\subsection{Mean and gaussian curvatures}

In order to extract some geometric properties of the surfaces $\mathbf{X}$ defined as in (\ref{defsurfaces}), we introduce a scalar product on the Lie algebra $su(n)$ given as
\begin{equation}
\langle A,B\rangle=\frac12\hbox{Tr}(AB),\quad A,B \in su(n).
\end{equation}
The first fundamental form [10] of the surface $\mathbf{X}$ is given by
\begin{equation}
\mathbf{I}=\langle d\mathbf{X},d\mathbf{X}\rangle=g_{++}dx_+^2+2g_{+-}dx_+dx_-+g_{--}dx_-^2,
\end{equation}
where $g_{\mu\nu}$ are the components of the metric tensor and are given explicitly by
\begin{equation}
g_{\pm\pm}=\langle\partial_{\pm}\mathbf{X},\partial_{\pm}\mathbf{X}\rangle=-\langle\partial_{\pm}\mathbb{P},\partial_{\pm}\mathbb{P}\rangle,\quad g_{\pm\mp}=\langle\partial_{\pm}\mathbf{X},\partial_{\mp}\mathbf{X}\rangle=\langle\partial_{\pm}\mathbb{P},\partial_{\mp}\mathbb{P}\rangle.
\end{equation}
Since, we are interested in solutions $\mathbb{P}$ given as in (\ref{solGmn}), we can show that our surfaces are conformal maps and that the metric components are as
\begin{equation}
g_{\pm\pm}=0,\quad g_{+-}=g_{-+}=\frac12\hbox{Tr}(\partial_+\mathbb{P}\partial_-\mathbb{P}).
\end{equation}
Note that the expression for $g_{+-}$ is identical to the expression of the Lagrangian density (\ref{ELproj}). Using the Brioschi formula [10], we may deduce the gaussian curvature $\mathcal{K}$ of the surface $\mathbf{X}$ associated to the solution $\mathbb{P}$:
\begin{equation}
\mathcal{K}=-\frac{1}{g_{+-}}\partial_+\partial_-\ln g_{+-}.
\end{equation}

Let us now calculate the expression for the mean curvature $\mathcal{H}$ [10] associated to solutions of the model. As we have shown, the considered surfaces are conformal maps and we know that the expression for the mean curvature is given as
\begin{equation}
\mathcal{H}=\frac12\hbox{Tr}(\mathbf{II}(\mathbf{I}^{-1})),
\end{equation}
where $\mathbf{II}$ is the second fundamental form defined as
\begin{equation}
\mathbf{II}=\langle\partial_+^2\mathbf{X},N\rangle dx_+^2+2\langle\partial_+\partial_-\mathbf{X},N\rangle dx_+dx_-+\langle\partial_-^2\mathbf{X},N\rangle dx_-^2=-\langle d\mathbf{X},d N\rangle.
\end{equation}
In the above expression, $N$ is a normal unit vector to the surface $\mathbf{X}$ and thus satisfies $\langle d\mathbf{X},N\rangle=0$. Using the conformal property of the surfaces, the mean curvature is given as
\begin{equation}
\mathcal{H}=\frac{\langle \partial_+\partial_-\mathbf{X},N\rangle}{g_{+-}}.
\end{equation}
 Due to the expression (\ref{derX}) and from the Euler-Lagrange equations (\ref{ELproj}), we easily get : 
\begin{equation}
\partial_+\partial_-\mathbf{X}=[\partial_+\mathbb{P},\partial_-\mathbb{P}]=[\partial_+\mathbf{X},\partial_-\mathbf{X}].
\end{equation}
We may thus define a unit normal vector $N$ to the surface $\mathbf{X}$ as
\begin{equation}
N=\frac{[\partial_+\mathbf{X},\partial_-\mathbf{X}]}{\|[\partial_+\mathbf{X},\partial_-\mathbf{X}]\|}=\frac{\partial_+\partial_-\mathbf{X}}{\|\partial_+\partial_-\mathbf{X}\|}\in su(n)
\end{equation}
remembering that $\langle d\mathbf{X},N\rangle=0$. Hence, we see that the mean curvature $\mathcal{H}$ is given by
\begin{equation}
\mathcal{H}=\frac{\|\partial_+\partial_-\mathbf{X}\|}{g_{+-}}=2\frac{\|[\partial_+\mathbb{P},\partial_-\mathbb{P}]\|}{\hbox{Tr}(\partial_+\mathbb{P}\partial_-\mathbb{P})}.\label{meancurv}
\end{equation}

\section{Solutions of $G(m,n)$, their associated surfaces and geometric characteristics}

\subsection{Special case of $\mathbb{C}P^{n-1}$ and the Veronese sequence}

In the late 80's, Bolton and \textit{al.} [9] fully classified constant gaussian curvature surfaces $\mathbf{X}$ associated to the solutions (\ref{solG1n}) of the $\mathbb{C}P^{n-1}$ model. Indeed, the set of all solutions is obtained from  the Veronese holomorphic curve $f$ defined as
\begin{equation}
f(x_+)=\left(1,\sqrt{\binom{n-1}{1}}x_+,\cdots,\sqrt{\binom{n-1}{r}}x_+^r,\cdots, x_+^{n-1}\right)^T.\label{Veronese}
\end{equation}
This holomorphic curve satisfies the following identity, which will prove to be useful in the rest of the paper,
\begin{equation}
\frac{\vert P_+^if\vert^2}{\vert P_+^{i-1}f\vert^2}=\frac{\alpha_{i,n}}{(1+\vert x\vert^2)^2}
\end{equation}
with
\begin{equation}
\alpha_{i,n}=i(n-i),\quad i=0,1,\cdots,n.
\end{equation}
In this case the gaussian curvature $\mathcal{K}$ takes the form
\begin{equation}
\mathcal{K}(Z_i)=\frac{4}{r_i(1,n)},\quad r_i(1,n)=r_i=n-1+2i(n-1-i).\label{gaussG1n}
\end{equation}
We see that the quantity $r_i$ admits an obvious symmetry given by
\begin{equation}
r_i=r_{n-1-i},
\end{equation}
which shows that some surfaces associated to non-equivalent solutions $Z_i$ and $Z_{n-1-i}$ have the same value of constant gaussian curvature. Note that when $n$ is odd, we omit $i= \frac{n-1}{2}$ since $Z_i=Z_{n-1-i}$. 

Thus we need an other geometric quantity to differentiate between them. In this case, the topological density is sufficient since we have
\begin{equation}
\mathcal{Q}(Z_i)=\frac{q_i(1,n)}{2(1+\vert x\vert^2)^2},\quad q_i(1,n)=q_i=n-1-2i,\label{topoG1n}
\end{equation}
where $q_i$ satisfy the relation:
\begin{equation}
q_{n-1-i}=-q_i.
\end{equation} 

To go further, we may express the quantity $\alpha_{i,n}$ in terms of the $r$'s and $q$'s given respectively in (\ref{gaussG1n}) and (\ref{topoG1n}). We get
\begin{equation}
\alpha_{m+j,n}=\frac{1}{2}\left[r_m+(2j-1)q_m\right]-j(j-1),\label{recurrencea}
\end{equation}
which will help us to express the geometric expressions in terms of the $r$'s and the $q$'s. As an example, we have 
\begin{equation}
\alpha_{m,n}=\frac12\left[r_m-q_m\right],\quad \alpha_{m+1,n}=\frac{1}{2}\left[r_m+q_m\right].
\end{equation}

The mean curvature $\mathcal{H}_i$, associated to the solution $Z_i$, is constant and is given by 
\begin{equation}
\mathcal{H}_i=2\frac{\sqrt{\alpha_{i,n}^2-\alpha_{i,n}\alpha_{i+1,n}+\alpha_{i+1,n}^2}}{\alpha_{i,n}+\alpha_{i+1,n}}=\frac{\sqrt{r^2_{i}+3q_{i}^2}}{r_{i}}.
\end{equation}
This geometric quantity is not necessary to differentiate surfaces of $\mathbb{C}P^{n-1}$ and of equal gaussian curvature, but will become relevant in higher dimensional grassmannians.


\subsection{Geometric characteristics of the Veronese curves of $G(m,n)$}

For the general grassmannian $G(m,n)$, we take the solution $\mathbb{P}_{\beta}$ given in (\ref{solGmn}) with $f$ chosen as for the Veronese holomorphic curve (\ref{Veronese}). The Lagrangian and topological densities are thus given by
\begin{equation}
\mathcal{L}(\mathbb{P}_{\beta})=\frac{r_{\beta}(m,n)}{2(1+\vert x\vert^2)^2},\quad 
\mathcal{Q}(\mathbb{P}_{\beta})=\frac{q_{\beta}(m,n)}{2(1+\vert x\vert^2)^2},
\end{equation}
where
\begin{equation}
r_{\beta}(m,n)=\sum_{j=1}^{n-1}(\beta_{j-1}-\beta_j)^2\alpha_{j,n},\quad q_{\beta}(m,n)=\sum_{j=1}^{n-1}(\beta_{j-1}-\beta_j)\alpha_{j,n}.\label{generalexprrq}
\end{equation}
Due to the fact that $\beta_j$ is 0 or 1, we easily deduce the following expressions:
\begin{eqnarray}
r_{\beta}(m,n)-q_{\beta}(m,n)&=&2\sum_{j=1}^{n-1}\beta_j\alpha_{j,n}-2\sum_{j=1}^{n-1}\beta_j\beta_{j-1}\alpha_{j,n}, \label{randq1}\\
r_{\beta}(m,n)+q_{\beta}(m,n)&=&2\sum_{j=1}^{n-1}\beta_{j-1}\alpha_{j,n}-2\sum_{j=1}^{n-1}\beta_j\beta_{j-1}\alpha_{j,n}.
\label{randq}
\end{eqnarray}
These expressions are slightly different from the ones obtained in the $\mathbb{C}P^{n-1}$ case. Indeed, they exhibit interaction between consecutive projectors $\mathbb{P}_{j-1}$ and $\mathbb{P}_{j}$ in the general expression of $\mathbb{P}_{\beta}$, which is naturally absent in the $\mathbb{C}P^{n-1}$ case.

Let us recall that the gaussian curvature of the surface associated to the solution $\mathbb{P}_{\beta}$ is given by
\begin{equation}
\mathcal{K}=\frac{4}{r_{\beta}(m,n)}.
\end{equation}
Furthermore, we may calculate the numerator of the mean curvature expression given as in (\ref{meancurv}), and we get
\begin{eqnarray}
\notag\|[\partial_+\mathbb{P}_{\beta},\partial_-\mathbb{P}_{\beta}]\|^2&\propto&\sum_{j=1}^{n-1}(\beta_{j-1}-\beta_{j})^2\alpha_{j,n}((\beta_{j-1}-\beta_{j})^2\alpha_{j,n}-\frac12(\beta_{j}-\beta_{j+1})^2\alpha_{j+1,n}\\&-&\frac12(\beta_{j-2}-\beta_{j-1})^2\alpha_{j-1,n}).
\end{eqnarray}
 For example, holomorphic solutions in the $G(m,n)$ case which are described by $\beta_i=1$ for $i=0,1,\cdots,m-1$ and $\beta_i=0$ for $i=m,m+1,\cdots,n-1$ lead to
 \begin{equation}
 r_{\beta}^{(hol)}(m,n)=q_{\beta}^{(hol)}(m,n)=\alpha_{m,n}=m(n-m),\quad \mathcal{H}_{\beta}^{(hol)}(m,n)=2.
 \end{equation}

Let us illustrate, in the following subsection of the case of the $G(2,n)$ system, the need to find the characteristics of distinguishing surfaces with the same gaussian curvature.

\subsubsection{The case of $G(2,n)$: some examples}

In this case, we have already computed the gaussian curvature for some surfaces associated to non holomorphic solutions of $G(2,n)$. For example, we have found [2] that
\begin{equation}
r_{2,3}(2,7)=r_{0,5}(2,7)=22.
\end{equation}
Now, if we compute the topological charge we find that
\begin{equation}
q_{2,3}(2,7)=q_{0,5}(2,7)=2.
\end{equation}
This means that we need another quantity to differentiate the geometry of these two surfaces.

The explicit forms of the quantities $r_{\beta}(2,n)$ and $q_{\beta}(2,n)$ are obtained directly from (\ref{randq1}) and (\ref{randq}). Indeed, we have to distinguish two cases: when $\beta_j=\beta_{j+1}=1$ (interaction), we get
\begin{eqnarray}
q_{j,j+1}(2,n)&=&2(n-2-2j), \\
 r_{j,j+1}(2,n)&=&2(n-2+j(n-2-j))=q_{j,j+1}(2,n)+2 \alpha_j;
\end{eqnarray} while
when $\beta_j=\beta_{k}=1$ for $k>j+1$ (absence of interaction), we get
\begin{eqnarray}
q_{j,k}(2,n)&=&2(n-1-j-k), \\
r_{j,k}(2,n)&=&2(n-1+j(n-1-j)+k(n-1-k))=q_{j,k}(2,n)+2 \alpha_j+2 \alpha_k.
\end{eqnarray}


Finally, we have (note that, for simplicity, we have set $\alpha_i=\alpha_{i,n}$)
\begin{eqnarray}
\mathcal{H}_{i,i+1}&=&2\frac{\sqrt{\alpha_{i}^2+\alpha_{i+2}^2}}{\alpha_{i}+\alpha_{i+2}},\\
\mathcal{H}_{i,i+2}&=&2\frac{\sqrt{\alpha_{i}^2-\alpha_{i}\alpha_{i+1}+\alpha_{i+1}^2-\alpha_{i+1}\alpha_{i+2}+\alpha_{i+2}^2-\alpha_{i+2}\alpha_{i+3}+\alpha_{i+3}^2}}{\alpha_{i}+\alpha_{i+1}+\alpha_{i+2}+\alpha_{i+3}},\\
\mathcal{H}_{i,j>i+2}&=&2\frac{\sqrt{\alpha_{i}^2-\alpha_{i}\alpha_{i+1}+\alpha_{i+1}^2+\alpha_{j}^2-\alpha_{j}\alpha_{j+1}+\alpha_{j+1}^2}}{\alpha_{i}+\alpha_{i+1}+\alpha_{j}+\alpha_{j+1}}.
\end{eqnarray}
The mean curvature is constant in each case and it helps to differentiate between the surfaces associated to $\mathbb{P}_{2,3}$ and $\mathbb{P}_{0,5}$ in $G(2,7)$. Indeed, we have
\begin{equation}
\mathcal{H}_{2,3}=\frac{2\sqrt{61}}{11},\quad \mathcal{H}_{0,5}=\frac{4\sqrt{7}}{11}.
\end{equation}
In Table 1, we give the examples of $G(2,n)$ with $n=4,5,6$.

\begin{table}[h]
\caption{The $G(2,4)$, $G(2,5)$ and $G(2,6)$ models}
\begin{center}
\begin{tabular}{|c|c|c|c|}
\hline
$(i,j)$&$r_{i,j}$&$q_{i,j}$&$\mathcal{H}_{i,j}$\\
\hline
$(0,1)$&4&4&2\\
\hline
$(1,2)$&6&0&$\sqrt{2}$\\
\hline
$(0,2)$&10&2&$\sqrt{\frac{2}{5}}$\\
\hline
\end{tabular}\quad \begin{tabular}{|c|c|c|c|}
\hline
$(i,j)$&$r_{i,j}$&$q_{i,j}$&$\mathcal{H}_{i,j}$\\
\hline
$(0,1)$&6&6&2\\
\hline
$(1,2)$&10&2&$\frac{2\sqrt{13}}{5}$\\
\hline
$(0,2)$&16&4&$\frac{\sqrt{7}}{4}$\\
\hline
$(0,3)$&14&2&$\frac{2\sqrt{11}}{7}$\\
\hline
$(0,4)$&8&0&$\sqrt{2}$\\
\hline
$(1,3)$&20&0&$\frac{1}{\sqrt{5}}$\\
\hline
\end{tabular}\quad \begin{tabular}{|c|c|c|c|}
\hline
$(i,j)$&$r_{i,j}$&$q_{i,j}$&$\mathcal{H}_{i,j}$\\
\hline
$(0,1)$&8&8&2\\
\hline
$(1,2)$&14&4&$\frac{\sqrt{106}}{7}$\\
\hline
$(2,3)$&16&0&$\sqrt{2}$\\
\hline
$(0,2)$&22&6&$\frac{\sqrt{58}}{11}$\\
\hline
$(0,3)$&22&4&$\frac{7\sqrt{2}}{11}$\\
\hline
$(0,4)$&18&2&$\frac{\sqrt{74}}{9}$\\
\hline
$(0,5)$&10&0&$\sqrt{2}$\\
\hline
$(1,3)$&30&2&$\frac{\sqrt{2}}{3}$\\
\hline
$(1,4)$&26&0&$\frac{7\sqrt{2}}{13}$\\
\hline
\end{tabular}


\end{center}
\end{table}

\subsubsection{The case of $G(2,n)$: some general results}

In this section, we give partial answers to the following question: If $q_{i,j}=q_{k,l}$ and $r_{i,j}=r_{k,l}$, is the mean curvature sufficient to differentiate the surfaces associated to $\mathbb{P}_{i,j}$ and $\mathbb{P}_{k,l}$?

In order to answer this question, we break the question into three steps.

 The first step is when $j=i+1$ and $l=k+1$. In this case, $q_{i,i+1}=q_{k,k+1}$ leads to $k=i$ and thus we are dealing with the same solution. We may exclude this case.
 
 The second step is the one when $j>i+1$ and $l>k+1$. In this case, the condition $q_{i,j}=q_{k,l}$ leads to $i+j=k+l$ which is equivalent to $l_{k,i}=i+j-k$. Then one wants $r_{ij}=r_{kl}$ which implies that 
 \begin{equation}
 r_{i,j}-r_{k,l}=r_{i,j}-r_{k,i+j-k}=4(i-k)(j-k)=0\quad \iff\quad i=k\quad \hbox{or}\quad j=k.
 \end{equation}
 But if $i=k$, then $j=l$ and we are dealing with the same solution. In the case when $j=k$, we have $i=l$, but this contradicts the original assumption that $j>i+1$. So we may exclude this case too.

 The third step is the one we have encountered for different grids: $j=i+1$ and $l>k+1$. The constraint $q_{i,i+1}=q_{k,l}$ leads to $l=2i-k+1$ and thus putting $l>k+1$, we get $i>k$. Now solving the constraint $r_{i,i+1}=r_{k,l}$, we get an expression for the dimension $n$ which is given by
 \begin{equation}
 n_{k,i}=3i+1-4k+\frac{2k(1+k)}{1+i},\quad i>k.
 \end{equation}

 So let us now look for a couple $(k,i)$ such that $n_{ki}$ is an integer. Once we have found such a couple, we must check that it satisfies  the constraints $l=2i-k+1<n$ and $i<n-1$.

We have 
 \begin{itemize}
 \item $n_{0,i}=3i+1\in\mathbb{N},\quad i>0$: This shows that the projectors $\mathbb{P}_{i,i+1}$ and $\mathbb{P}_{0,2i+1}$ have the same gaussian curvature and the topological charge. But can the corresponding mean curvatures be different? If they are the same we have
 \begin{equation}
 \left(\frac{\mathcal{H}_{0,2i+1}}{\mathcal{H}_{i,i+1}}\right)^2_{n=3i+1}=\frac{2+i+3i^2+i^3+2i^4}{2-6i+i^2+8i^3+4i^4}=1\quad \iff\quad i=1.
 \end{equation}
 But the case $i=1$ can be easily understood since in this case $n_{0,1}=4$ and we are thus comparing the projectors $\mathbb{P}_{1,2}$ and $\mathbb{P}_{0,3}$ which correspond to the same solution of the $G(2,4)$ model. So we see that in this case the mean curvatures are different.
 \item $n_{1,i}=-3+3i+\frac{4}{1+i}\in\mathbb{N},\quad i>1$: This shows that the only admissible $i$ is $i=3$. In this case, we have $n_{1,3}=7$ and we are comparing the projectors $\mathbb{P}_{3,4}$ and $\mathbb{P}_{1,6}$. Using the completeness relation, this is the same as comparing the projectors $\mathbb{P}_{2,3}$ and $\mathbb{P}_{0,5}$ which is consistent with the properties of the grid of $G(2,7)$. As we already know, the mean curvatures in this case are different.
 \item $n_{2,i}=-7+3i+\frac{12}{1+i}\in\mathbb{N},\quad i>2$: Putting all the constraints together, this shows that the only admissible $i$ are $i=11$ and $i=5$. In the first case, we get $n_{2,11}=27$ and we are comparing the projectors $\mathbb{P}_{11,12}$ and $\mathbb{P}_{2,21}$. The second possibility leads to $n_{2,5}=10$ and we are then comparing the projectors $\mathbb{P}_{5,6}$ and $\mathbb{P}_{2,9}$. As before, we can show that the mean curvatures are different.
 \end{itemize}

Let us make some further comments. Indeed, if we set $l=n-1$, then we have that $n=2i+2-k$ which can then be compared with the formula for $n_{k,i}$. Doing so, we obtain $i=k-1$ or $i=1+2k$. The case $i=k-1$ must be rejected due to the constraint that $i>k$. This means that we are left with the unique solution $i=1+2k$. We thus have $n_{k,1+2k}=4+3k$ and we are comparing the surfaces associated to the projectors $\mathbb{P}_{1+2k,2+2k}$ and $\mathbb{P}_{k,3+3k}$. This general result is consistent with the above examples. Furthermore, we can show that
\begin{equation}
\left(\frac{\mathcal{H}_{k,3+3k}^{(4+3k)}}{\mathcal{H}_{1+2k,2+2k}^{(4+3k)}}\right)^2=\frac{9+18k+18k^2+9k^3+2k^4}{9+36k+49k^2+24k^3+4k^4}=1\quad \iff\quad k=-\frac92, -2, -1, 0.
\end{equation}
Since $k\geq 0$, the mean curvatures are different. 

In Table 2, we give some possible values of $i$.
\begin{table}[h]
\caption{Some possible values of $i$ in the expression for $n_{k,i}$}
\begin{center}
\begin{tabular}{|c|c|c|c|}
\hline
$i$&$n_{k,i}$&$l_{k,i}$&$\hbox{Comments}$\\
\hline
$2k(1+k)-1$&$6k^2+2k-1$&$4k^2+3k-1$&$k>0$\\
\hline
$k(1+k)-1$&$k(3k-1)$&$2k^2+k-1$&$k>1$\\
\hline
$2k+1$&$3k+4$&$3(k+1)$&$k\geq 0$\\
\hline
$2k-1$&$3k-1$&$3k-1$&$\hbox{N/A}$\\
\hline
\end{tabular}
\end{center}
\end{table}
This table contains all the examples mentioned above. The above discussion works for $k=3,4$, but for $k=5$ new cases arise. Indeed, we have two new cases: $n_{5,19}=41$ with $l_{5,19}=34$ and $n_{5,14}=27$ with $l_{5,14}=24$. The later case can be explained using the fact that $2k(1+k)$ in the expression of $n_{k,i}$ is always divisible by 4. So we have summarized all this information in Table 3.
\begin{table}[h]
\caption{Values of $i$ for which $2k(1+k)$ is divisible by 4}
\begin{center}
\begin{tabular}{|c|c|c|c|}
\hline
$i$&$n_{k,i}$&$l_{k,i}$&$\hbox{Comments}$\\
\hline
$m(2m+3)$&$6m^2+m+1$&$4m(m+1)$&$k=2m+1\quad m\geq 2$\\
\hline
$m(2m+1)-1$&$6m^2-5m+2$&$4m^2-1$&$k=2m\quad m\geq 2$\\
\hline
\end{tabular}
\end{center}
\end{table}
We conjecture that, that as $k$ increases, the number of cases will increase and it is totally dependent on the prime decomposition of the factor $2k(1+k)$ in the expression for $n_{k,i}$.

\subsection{Non-Veronese holomorphic solutions}
We have conjectured [1] that we can construct a holomorphic solution in $G(m,n)$ of constant gaussian curvature $\mathcal{K}=\frac{4}{r}$ for all integer values of $r=1,2,\cdots,\alpha_{m,n}$. The maximal value $r=\alpha_{m,n}=m(n-m)$ was obtained from the Veronese holomorphic curve (\ref{Veronese}) and its $m-1$ consecutive derivatives. The values of $r=1,2,\cdots, m(n-1-m)$ are obtained from the natural immersion of $G(m,n-1)$ into $G(m,n)$. The other values are not obtained from such immersions nor from the Veronese curve. Furthermore, for the same value of the missing $r$, we may find non-equivalent solutions $Z_{i}=\hat{Z}_{i}\hat{L}_{i}, \ i=1,2$. As an example, in the $G(2,5)$ case, we have obtained [1] two non-equivalent holomorphic solutions corresponding to $r=5$ parametrized by
\begin{equation}
\hat{Z}_1^T=\left(\begin{array}{ccccc}
1&0&\sqrt{5}x_+&\sqrt{5}x_+^2&0\\
0&1&\sqrt{5}x^2_+&\frac{7}{\sqrt{5}}x_+^3&\frac{1}{\sqrt{5}}x_+^3
\end{array}\right),\quad \hat{Z}_2^T=\left(\begin{array}{ccccc}
1&0&x_+&\frac{1}{\sqrt{5}}x_+^2&0\\
0&1&2x_+&\frac{7}{\sqrt{5}}x_+^2&\sqrt{5}x_+^3
\end{array}\right).
\end{equation}
Of course, we know that in the holomorphic case the Lagrangian density corresponds to the topological one. Since the Lagrangian density of these two solutions is the same this is also the case for the topological density. Hence, the mean curvature will make the difference. Indeed, using $Z=\hat{Z}\hat{L}$ that satisfies $\hat{Z}^{\dagger}\hat{Z}=(\hat{L}\hat{L}^{\dagger})^{-1}$, we get
\begin{equation}
\mathbb{P}=\hat{Z}(\hat{Z}^{\dagger}\hat{Z})^{-1}\hat{Z}^{\dagger}.
\end{equation} 
In this case, the mean curvature is not a constant, ie $\mathcal{H}_i=\mathcal{H}_i(\vert x\vert^2)$ . Morevover,  we can show that $\mathcal{H}_1\neq \mathcal{H}_2$. Indeed from (\ref{meancurv}), we get
\begin{equation*}
\left(\frac{\mathcal{H}_1}{\mathcal{H}_2}\right)^2=\frac{\mathcal{P}_1(\vert x\vert^2)}{\mathcal{P}_2(\vert x\vert^2)},\quad \mathcal{P}_2(y)=y^6\mathcal{P}_1\left(\frac{1}{y}\right)=25+110y+285y^2+428y^3+355 y^4+150y^5+25 y^6.
\end{equation*}

\section{Conclusion}

In this paper, we have studied the geometric properties of surfaces constructed from the solutions of the two-dimensional $G(m,n)$ sigma model. The aim of our work was to see whether we can differentiate surfaces of equal constant gaussian curvature by involving also the topological density and the mean curvature. This problem originated from the complete classification of constant gaussian curvature surfaces [9] associated to Veronese holomorphic curves for $\mathbb{C}P^{n-1}$. In this classification, some non-equivalent solutions had  identical gaussian curvature due to the symmetry property of the gaussian curvature. In these cases the topological charges had different values.

We have not fully solved the problem which has turned out to be more complicated than originally envisaged. So here we report 
where we are at the moment. 
We have  generalized the previous results to more general grassmannians. We have found explicit expressions for the gaussian curvature, the topological density and the mean curvature for these solutions. We have shown, in the case of $G(2,n)$, that some non-equivalent solutions may have identical gaussian curvature but also identical topological densities. This has led us to consider the second fundamental form of these surfaces and  we have computed their mean curvature. Some partial results for the $G(2,n)$ case show that the mean curvature is sufficient to distinguish surfaces with identical gaussian curvature and topological charge. Furthermore, we have shown that the projectors having this property are all of the form $\mathbb{P}_{i,i+1}$ and $\mathbb{P}_{k,l}$ with $l>k+1$.

The case of holomorphic solutions which are not of the Veronese type and that can not be obtained from immersions of lower dimensional grassmannians is really challenging since we have no general formula for these cases. Our results are complete for  the $G(2,5)$ model where we have shown that all solutions of equal gaussian curvature and topological density have distinct mean curvature.

\section*{Acknowledgments}
This work has been supported in part by research grants from Natural sciences and engineering research council of Canada (NSERC). Laurent Delisle also acknowledges a NSERC Postdoctoral fellowship.

\section*{References}

\end{document}